\documentclass[aps, pre, twocolumn, floatfix]{revtex4-1}
\usepackage{blindtext}
\usepackage{graphicx}
\usepackage{graphics}
\usepackage{bm}   % for math
\usepackage{verbatim}   % for math
\usepackage{amsfonts}
\usepackage{amsmath}
\usepackage{bm}
\usepackage{amssymb}
\usepackage{subfigure}
\usepackage{adjustbox}
\usepackage{lipsum}

\usepackage{adjustbox}

\usepackage{stfloats}

\usepackage{amsmath,amscd}
\usepackage {extarrows}
\usepackage{algorithm,algpseudocode}
	
\makeindex

\begin{document}

\title {Creep response of athermal amorphous solids under imposed shear stress}

\author{ Suman Dutta}
\affiliation{International Centre for Theoretical Sciences - Tata Institute of Fundamental Research\\
Survey No 51, Hessaraghatta Hobli, Shivakote, Bangalore 560089, India}

\author{Kirsten Martens}
\affiliation{Universit\'e Grenoble Alpes, CNRS, LIPhy, 38000 Grenoble, France}

\author{Pinaki Chaudhuri}
\affiliation{The Institute of Mathematical Sciences, Taramani, Chennai 600113, India}

%\email{Correspondence to: TBA}

%\footnote {$^{\ddag}$ Contributed equally}

\date{~\today}

\begin{abstract}
Yield stress materials fail when the imposed stress crosses a critical threshold. A well-known dynamical response to the applied stress is the phenomenon of creep where the cumulative deformation grows sublinearly with time, prior to failure or arrest. Using extensive molecular dynamics simulations, we study such response for a model amorphous system, in the athermal limit, and probe how the annealing history of the initial state determines the observed behaviour to an applied shear stress. Further, we analyze the microscopic dynamics in the vicinity of the yield threshold, using large systems, and characterize the spatiotemporal signatures towards arrest or flow, at different scales.
\end{abstract}

\maketitle

\section{Introduction}

The mechanical properties of amorphous materials determine the occurrence of diverse natural phenomena (e.g. landslides, avalanches, etc.) as well a large number of products and applications (e.g in the form of glasses, gels, foams, emulsions, colloids etc.). Therefore, developing a physical understanding,  from a microscopic perspective, of the processes that lead to the observed mechanical response is necessary and hence draws a significant amount of research efforts utilizing a combination of experiments, numerical modelling at micro as well as meso scales, and analytical frameworks (e.g. see reviews \cite{schuh2007mechanical,rodney2011modeling, bonn2017yield, nicolas2018deformation}). The characteristic feature of amorphous solids, soft or hard, is the existence of a yield stress, exceeding which the material fails or enters steady flow. The approach to yielding occurs via the proliferation of local plastic events which self-organise to form extensive avalanches. Even after yielding, such spatially organised events persist, e.g. in the form of shear bands \cite{fielding2014shear}. It is now well-established that the transient response displayed by these materials depends on the preparation history of the amorphous state, and recent theoretical work has focused on probing this in terms of ductile/brittle response depending upon the degree of annealing \cite{ozawa2018random, popovic2018elastoplastic, barlow2020ductile, richardfinite, richardbrittle2021, rossi2022finite, pollard2022yielding}.
%One of the major goals in this context, is to develop a unique theoretical framework {which describes} transient phenomena like stress overshoots prior to yielding (e.g.~in metallic glasses
%\cite{kawamura1999stress, lu2003deformation, maass2012shear} 
%and soft materials 
%\cite{amann2013overshoots, divoux2011stress, dzuy1983yield, rogers2010time, zausch2008equilibrium, sentjabrskaja2014transient, koumakis2012yielding, koumakis2012direct}), delayed failure in creep experiments \cite{chaudhuri2013onset, sentjabrskaja2015creep, cipelletti2020microscopic}, along with steady state properties, e.g. strongly non-linear flow curves \cite{agoritsas2017non}.

One of the typical mechanical protocols involves the application of an externally imposed step stress on the material and in response, amorphous solids display an interesting dynamical feature known as creep, wherein the time evolution of the emergent strain, resulting from the applied loading, increases sub-linearly \cite{betten2008creep}. In the case of thermal glasses or gels, the creep process can in principle proceed indefinitely with diverse sub-linear exponents reported in literature (e.g. see \cite{bauer2006collective, Divoux2011a,Leocmach2014,ballesta2016creep,chaudhuri2013onset,sentjabrskaja2015creep,landrum2016delayed, cipelletti2020microscopic,merabia2016thermally,joshi2018yield, cabriolu2019precursors,dijksman2022creep, bhowmik2022creep}) and a logarithmic growth has been proposed in the asymptotic regime \cite{siebenburger2012creep}. However, in the case of athermal or non-Brownian systems, the sub-linear growth can suddenly terminate in the form of a dynamical arrest, if the amount of applied stress is below the static yield stress threshold \cite{paredes2013rheology, bonn2017yield}. Above the threshold, the material eventually reaches a steady flow, as in the case of soft glassy materials, or can fail catastrophically, as in the case of brittle solids. Although there are several experimental studies of creep in non-Brownian materials, numerical or analytical investigations have been limited. Only recently, using mean field or elasto-plastic models, athermal creep has been analysed in some details \cite{PhysRevLett.120.028004, liu2018creep, liu2021elastoplastic, popovic2022scaling}. However, most of these studies provide the overall macroscopic analysis and the spatio-temporal aspects need to be explored in a systematic way to analyse the processes at play.

In this work, we study how a two-dimensional dense athermal amorphous solid responds to applied shear stress, using extensive molecular dynamics simulations. Our study is two-fold. The initial part focuses on demonstrating how the annealing history for the preparation of the athermal solid determines the macroscopic response, specially the transient regime, including the determination of the static yield threshold. In the second half, we analyse the mechanism of dynamical arrest or flow, around the yield threshold, and reveal the spatio-temporal plasticity at work.
 
This manuscript is organised as follows. After the introductory discussion in Section I, we provide details of the model and the numerical methods in Section II. This is followed an extensive discussion of our findings in Section III. Finally, we have a concluding discussion in Section IV.

\section{Model and Methods}

In this study, we consider a model two-dimensional glass-forming binary mixture having two different sizes \cite{lanccon1988two, lanccon1986thermodynamical}, whose mechanical response has been extensively studied \cite{falk1998dynamics, patinet2016connecting, barbot2018local}. The ratio of such large (L) and small (S) is given by $\frac{N_{L}}{N_{S}}=\frac{1+\sqrt{5}}{4}$ where $N_{L}$ and $N_{S}$ are the number of L and S particles respectively. The interaction between any pair of particles is the following:
%separated at a distance $r_{i,j}=|{\bf r}_{i}-{\bf r}_{j}|$ is %$V(r_{i,j})$. 
%The interaction between any pair of particles having co-ordinates ${\bf r}_{i}$ and ${\bf r}_{j}$,  is given by
%
\begin{eqnarray}
V_{\alpha\beta}(r) &=&
4\epsilon_{\alpha\beta}\left[\left(\sigma_{\alpha\beta}/r\right)^{12}-
\left(\sigma_{\alpha\beta}/r\right)^{6}\right] \, .
\label{poteqn}
\end{eqnarray}
where $r=|{\bf r}_{i}-{\bf r}_{j}|$, and $\alpha, \beta$ correspond to the identities $S$ or $L$.
The values of the interaction parameters are set to $\epsilon_{\textrm{SL}}
= 1.0$, $\epsilon_{\textrm{SS}} = \epsilon_{\textrm{LL}} = 0.5\epsilon_{\textrm{LS}}$, $\sigma_{\textrm{LS}}=1$,
$\sigma_{\textrm{LL}} = 2\sin{\pi/5}$, $\sigma_{\textrm{ss}} = 2\sin{\pi/10}$. In the following,
we use $\epsilon_{\textrm{LS}}$ and $\sigma_{\textrm{LS}}$ as the
unit for energy and length, respectively.  The cutoff radius in
Eq.~(\ref{poteqn}) is chosen as $R_{c} = 2.5\sigma_{\textrm{LS}}$ and the potential is smoothened out near the cutoff \cite{patinet2019origin}.  As the time unit, we use $\sqrt{{m\sigma_{\textrm{LS}}^{2}} /
\epsilon_{\textrm{LS}}}$, where $m$ is the mass of a particle that
is considered to be equal for both type of particles, i.e.~$m =
m_{\textrm{L}} = m_{\textrm{S}} = 1.0$.  More details about the
model can be found in Ref.\cite{falk1998dynamics, patinet2016connecting, barbot2018local}.

%The form of $V(r_{i,j})$ is same as in Ref.\cite{}. 

The equation of motion for any particle within the athermal assembly is given by: 
\begin{equation}
	{m_{i}}\ddot{{\bf r}_{i}}= \sum^{N}_{j\ne i} {\bf f}^{int}_{(i,j)}+ {\bf f}^{D}_{(i,j)}
\end{equation}
Here, the total force acting on the particle is a sum over the interaction forces coming from the potential discussed above, viz. $ {\bf f}^{int}_{(i,j)}=- \vec{\nabla} V_{(i,j)}$, and a dissipative force, $\bf{f}^{D}_{(i,j)}=-\zeta w^{2}(r_{(i,j)}) ~ (\hat{{\bf r}}_{(i,j)} ~. ~{\bf v}_{(i,j)})\hat{{\bf r}}_{(i,j)}$ which stabilizes the system against runaway heating effects resulting from the continuously imposed external drive in the form of shear.
%and the random force, $\bf{f}^{R}_{(i,j)}=\sqrt{2k_{B}T}\zeta w(r_{(i,j)})\theta_{(i,j)}~\hat{{\bf r}_{(i,j)}}$. %The equations (1-2) ensure the correct thermostatting of the system by canceling the drifting velocities introduced due to shear and warrant Galilean-invariance and conservation of local momenta. 
%Note that for an athermal assembly, ${\bf f}^{R}=0$.
Apart from the equations of motion corresponding to the particle co-ordinates, we also have an equation of motion of the macroscopic shear-rate ($\dot{\gamma}$), resulting from the  constant shear stress ($\sigma_0$) imposed along the $xy$-plane. The constant stress protocol is implemented via a feedback control scheme \cite{vezirov2015manipulating, cabriolu2019precursors}, which leads to the following form for the time evolution of shear rate due to the imposed stress:
\begin{equation}
	{\ddot{\gamma}(t)}=B[\sigma_0-\sigma_{xy}(t)]
\end{equation}
where $\sigma_{xy}=[m_{i}(v^{x}_{i}(t)v^{y}_{j}(t))+\sum^{N}_{j>i}(r^{x}_{(i,j)} ~f^{y}_{(i,j)}(t))]/V$, is the Irving-Kirkwood expression for the shear stress, using the velocity components, $v^{x}$ and $v^{y}$ at single particle level. Here, we choose a suitable value of damping parameter, $B=1$ such that it can capture the correct long time dynamics and the fluidization of the material. Previously, as mentioned earlier, the rheological properties of this binay mixture model has been studied and the dynamical yield threshold has been reported to be $\sigma_d=0.5107$ \cite{liu2021elastoplastic} All imposed shear stress values ($\sigma_0$) in this work will be expressed in units of $\sigma_d$.
%We take 16 large samples of each of the inherent structures $HTL$, $ESL$ and $GQ$ and apply constant stresses ($\sigma$) varying the magnitude at $T=0$. 

The molecular dynamics simulations were done using LAMMPS \cite{plimpton1995fast}. The cosnidered system sizes are $N=102400, 409600$ which lead to respective box-lengths of $L=316.174, 632.348$ for the density at which we probe the properties of the model system. The time step for the numerical integration was chosen to be $\tau=0.001$. We prepare equilibrium thermal states at $T=2.98, 0.351$ and subsequently, we generate  inherent structures states, labelled $HTL$ and $ESL$, via energy minimization of states sample from the respective equilibrated ensembles. The third ensemble, labelled $GQ$, was prepared by slowly quenching the thermal sample at $T=0.351$ to $T=0.0299$ in $10^6 \tau$.   

%The rheological properties of this model has been studied previously and the dynamical yield threshold has been reported to be $\sigma_d=0.5107$. All imposed shear stress values in this work will be expressed in units of $\sigma_d$.

\begin{figure*}[t]
\centerline{\includegraphics[scale=0.6]{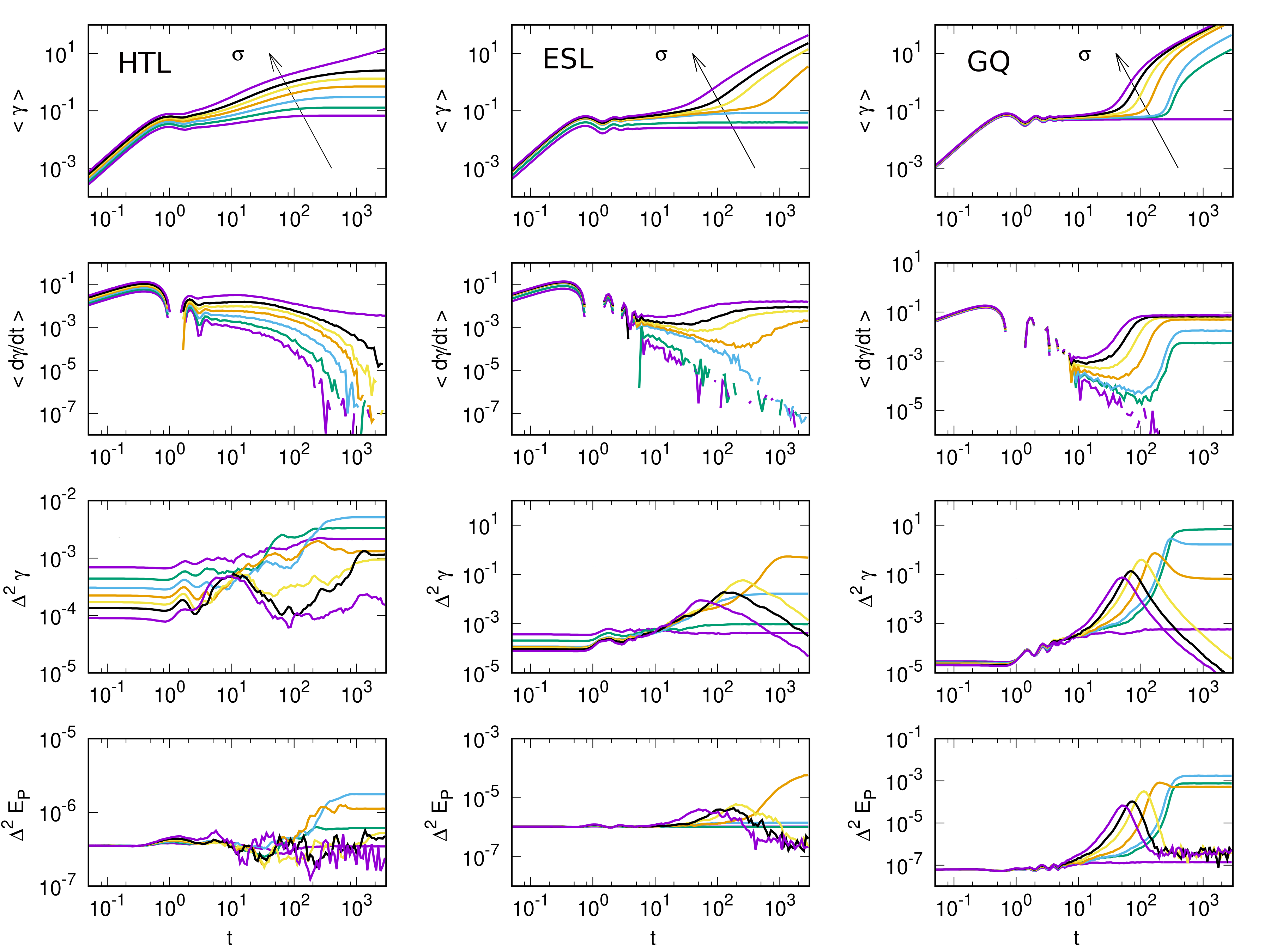}}
\caption{{\bf Response of states having different preparation histories}. $N=102400$. The three columns correspond to histories labelled as HTL, ESL, GQ (see text for details). In each column, we show time evolution of ensemble-averaged (a) macroscopic strain,  $\langle\gamma(t)\rangle$ (b) macroscopic shear-rate, strain-rate $\dot{\gamma}$ (labelled as  $\langle d\gamma(t)/dt\rangle$) (c) strain fluctuations, $\Delta^{2}\gamma(t)$ and (d) potential energy fluctuations, $\Delta^{2}E_p(t)$. For HTL states, data is shown for $\sigma_0/\sigma_d=0.39,0.489,0.587,0.685,0.783,0.881,1.077$. For ESL states, data is shown for  $\sigma_0/\sigma_d=0.587,0.783,1.038,1.077,1.126,1.175,1.273$. For GQ states, data is shown for $\sigma_0/\sigma_d=1.468,1.547,1.566,1.615,1.664,1.713,1.762$.}
	\label{fig1}
\end{figure*}

\begin{figure*}[t]
\centerline{\includegraphics[scale=0.45]{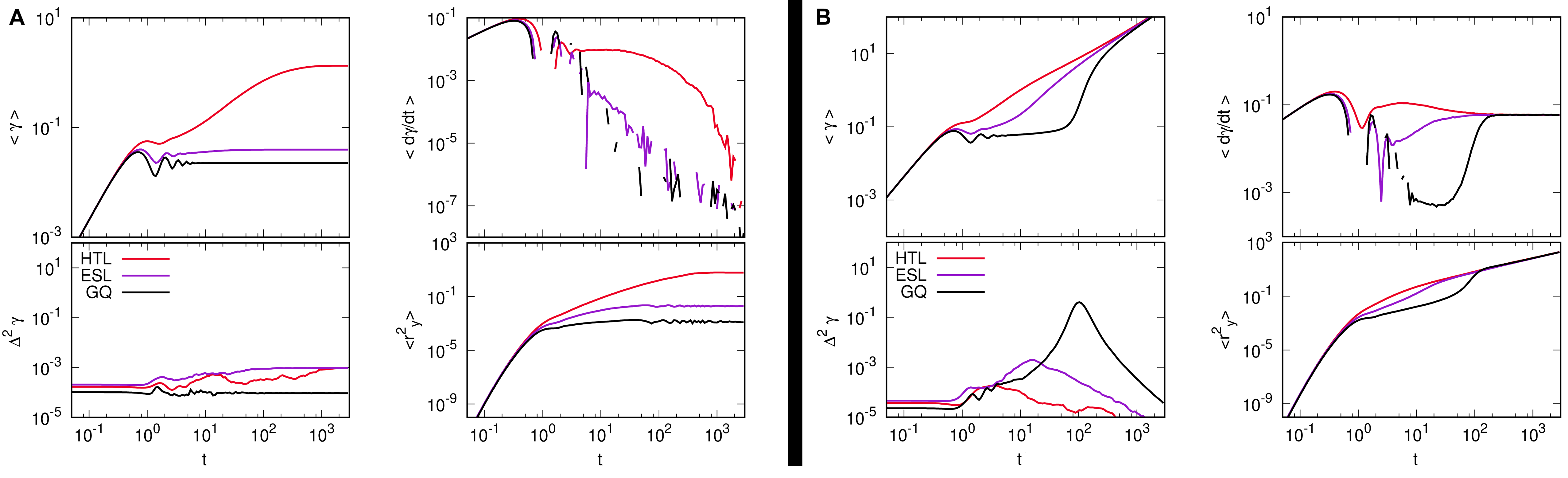}}
\caption{{\bf Comparison across different preparation histories labelled as HTL, ESL, GQ :} (A) For $\sigma_{0}/\sigma_d=0.783$, where eventually flow is arrested in all cases; and (B)  for $\sigma_{0}/\sigma_d=1.664$ where there is a long-time steady flow in all cases. In each case,  time evolution of (a) strain $\langle\gamma(t)\rangle$, (b) strain rate strain-rate $\dot{\gamma}$ (labelled as  $\langle{d\gamma/dt}\rangle$), (c) strain fluctuations $\Delta^{2}\gamma(t)$, and (d) MSD along gradient direction $\langle{r^{2}_{y^2}}\rangle$, are shown. The data is shown for a system size of $N=102400$, for the three different preparation protocols, in each case, as marked.}
\label{fig2}
\end{figure*}

\section{Results}

\subsection{Creep data: averages and fluctuations}

We start our discussion of results by first analysing the macroscopic response to applied shear stress. Since we are imposing a shear stress on the system, our observable of interest is the cumulative deformation that the system undergoes with time which is quantified via the shear strain, $\gamma(t)$, and the corresponding rate of deformation quantified via the shear-rate $\dot{\gamma}(t)$. When the rate of deformation, i.e.  $\dot{\gamma}(t)$, becomes constant, then $\gamma(t) \sim t$; this would correspond to a non-equilibrium steady state. If the system does not eventually flow but gets stuck, then the amount of deformation will remain fixed after that, i.e. $\gamma(t)$ reaches a plateau and $\dot{\gamma}(t)$ goes to zero. An intermediate regime is where the  $\gamma(t)$ grows sublinearly with time, which is termed as creep. To remind, below  an applied stress threshold, we expect the dynamics to get arrested, viz. $\dot{\gamma}(t)$ will vanish, and above the threshold, 
steady flow, i.e. a constant $\dot{\gamma}(t)$, is expected in compliance with the steady state macroscopic rheological flow curve.

In the first and second rows of Fig.\ref{fig1}, respectively, we show the data for $\gamma(t)$ and $\dot{\gamma}(t)$, averaged over the ensemble of states considered, with the three columns corresponding to the data for the three different preparation histories, viz. HTL, ESL, GQ. %To remind, the dynamical yield stress for the model is 0.5107. 
The first important finding is that the yield threshold, under applied stress, is dependent on  preparation history.  For states labelled HTL and ESL, within the resolution of our study, we observe flow for  $\sigma_0/\sigma_d \geq 1.077$, and for smaller values of  $\sigma_0$, the system gets stuck at long times. However, for GQ, the threshold changes dramatically; there is no flow for $\sigma_0/\sigma_d \leq 1.468$. Thus, with improved annealing, i.e. for more low lying inherent structures states on the energy landscape, the yield threshold shifts to higher values. This is consistent with findings in athermal quasistatic shear, where the peak stress during yield is observed to increase with increase annealing \cite{ozawa2018random}. Since the yielding stress threshold increases with preparation history, the flow rate  at which steady state is observed also changes; lower lying energy states have a larger onset flow rate, to be consistent with the underlying macroscopic flow curve.  

Note that the shape of the transient $\dot{\gamma}(t)$ have different trends for the three different histories. For HTL, $\dot{\gamma}(t)$ decreases with time, and then eventually vanishes or reaches a steady state. For the other two histories, $\dot{\gamma}(t)$ decreases in the case when the system gets arrested at long times. On the other hand, when there is eventual steady flow, $\dot{\gamma}(t)$ has a non-monotonic shape, with $\dot{\gamma}(t)$ initially increasing, goes through a minimum and then increases to the final steady state. Broadly, these features in $\dot{\gamma}(t)$ observed in our numerical simulations are consistent with experimental data \cite{divoux2011stress, siebenburger2012creep}.

Next we study the fluctuations in these macroscopic quantities, as measured across the different trajectories within the ensemble starting from independent amorphpus states. In this context,  while studying creep in thermal glasses, we have shown that the onset of large scale plasticity that leads to steady flow is demarcated by a peak in strain fluctuations across trajectories, and the timescale of the peak depends upon applied stress \cite{ cabriolu2019precursors}. These peaks are clearly visible for the GQ, ESL states; see third row in Fig.\ref{fig1} . Further, we now see that the fluctuations in the potential energy across trajectories also exhibit a similar behaviour, with the peak approximately occurring when the peak in strain fluctatuations occur; see fourth row in Fig.\ref{fig1}. This allows for using energy fluctuations as a simpler way to identify onset of plasticity. Note that the peak location is at a time later than the location of the minimum in $\dot{\gamma}(t)$.

In Fig.\ref{fig2}, we further analyse the nature of the transient behaviour, for the three different preparation histories, by focusing on the response to the same applied stress, viz. two different magnitudes, viz. $\sigma_0/\sigma_d=0.783, 1.664$. In the case of $\sigma_0/\sigma_d=0.783$, the deformation gets arrested  independent of history (see left panel of Fig.\ref{fig2}), whereas for $\sigma_0/\sigma_d=1.664$, there is a steady flow at long times in all the three cases (see right panel of Fig.\ref{fig2}). In each case, we show data for the time evolution of (a) strain $\langle\gamma(t)\rangle$, (b) strain rate strain-rate $\dot{\gamma}$, (c) strain fluctuations $\Delta^{2}\gamma(t)$, and (d) single particle mean squared displacement (MSD) measured along gradient direction $\langle{r^{2}_{y^2}}\rangle$ which is a measure of non-affine motion generated due to plasticity originating from the imposed shear. For the smaller stress value, the system gets arrested very quickly for the ESL and GQ states, whereas for the HTL states, there is a prolonged transient regime before the system gets arrested; this is evident from both the macroscopic strain data as well as the MSD data which shows a long time plateau indicating that the motion of the particles is frozen in time. The situation is very different for the larger applied stress, where there is a long time steady state with a well-defined constant $\dot{\gamma}$ for all the three cases and the MSD data also show identical long-time diffusive regime. We note that the more low lying inherent structure states require longer time to reach the steady-state, which is also known for the case of imposed shear rate (e.g. \cite{lamp2022brittle}).  Interestingly, only in this case, there is pronounced signal in the fluctuations data, with the heterogeneity across trajectories larger for the low-lying states.  Such non-monotonic behaviour in fluctuations is not present in the case of the smaller applied stress. Thus, this exercise also validates the use the fluctuations signal as a diagonistic for onset of large scale steady flow. 

%{\em Add snapshots of MSD for transient states?}

\subsection{Comparing paths to dynamical arrest and fluidization: a microscopic analysis}

After exploring the macroscopic response across the different ensembles corresponding to different preparation histories, we now focus on probing at a microscopic scale how fluidization proceeds in the vicinity of the estimated yield threshold. For this study, we consider only the low-lying GQ states, where the onset of plasticity is  dramatic; see the data of shear-rate in the third column  of Fig.\ref{fig1} where we observe that typically $\dot{\gamma}(t)$ initially decreases, goes through a minimum, before a sudden increase to reach the steady state.

\begin{figure*}[t]
\centerline{\includegraphics[scale=0.43]{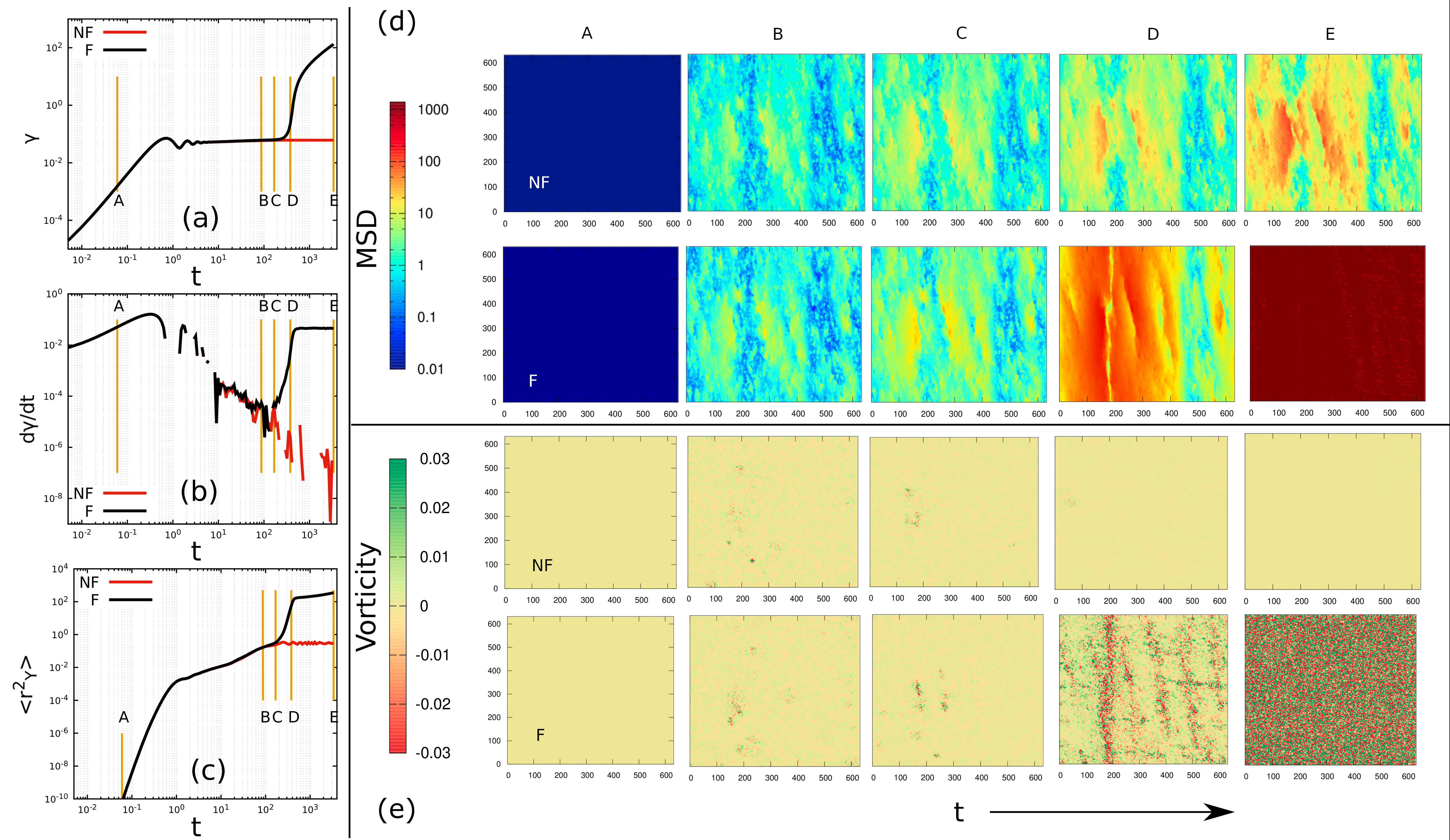}}
\caption{{\bf Comparative study of trajectories for $\sigma_0/\sigma_d=1.547$ and $\sigma_0/\sigma_d=1.557$, starting from same initial state.} $N=409600$.
Time evolution of (a) strain, $\gamma(t)$, (b) strain-rate $\dot{\gamma}$ (labelled as  $d\gamma/dt$), (c) mean-squared displacement, $r^2_y(t)$: trajectory for $\sigma_0/\sigma_d=1.557$ gets fluidized (F) whereas it gets arrested, i.e. non-fluidized  (NF) for $\sigma_0/\sigma_d=1.547$. (d) Local MSD maps (see text for details), using a coarse-graining scale of $\ell^2$, with $\ell=2.47$, for $\sigma_0/\sigma_d=1.547$ (first row; labelled NF) and $\sigma_0/\sigma_d=1.557$ (second row; labelled F). {The maps are shown at the time points $t= 0.06$ (A), $87.6$ (B), $167.7$ (C), $377.3$ (D) and $3369.7$ (E), as marked in (a), (b) (c).} (e) Local vorticity maps (see text for details), using coarse-graining scale used in (d), computed at the same time points for both NF (first row) and F (second row) trajectories.}
\label{fig3}
\end{figure*}

In particular, we discuss the following thought experiment where we start with the same initial state and follow the trajectories corresponding to two close-by applied stress values of $\sigma_0/\sigma_d=1.547, 1.557$; see Fig.\ref{fig3}(a)-(c) for data on strain, strain-rate and MSD. We can clearly see that for this initial state and applied shear stress of $\sigma_0/\sigma_d=1.547$, the single particle dynamics (quantified via MSD) gets arrested at long times and thereby the macroscopic deformation also stops; labelled NF. However, for $\sigma_0/\sigma_d= 1.557$, the system flows with a constant shear-rate at long times; labelled F.  Thus, with a very small increase in the magnitude of the applied shear stress, the fate of the system, starting from the same initial state, changes from being arrested at long times to that of a steady flow. However, the interesting and significant feature is that the evolution of strain, shear-rate, MSD with time are very similar for both cases, till around $10^2$, after which the trajectories diverge. If one follows $\dot{\gamma}(t)$, then beyond this point of divergence, for $\sigma_0/\sigma_d= 1.557$, there is suddenly a rapid increase and the system enters a steady flow regime whereas for $\sigma_0/\sigma_d=1.547$, it continues to decrease and eventually vanishes to become arrested, i.e. the system gets absorbed in a local minima of its underlying potential energy landscape. The MSD data also show similar signatures -- for the smaller stress, it evolves to the asymptotic plateau, whereas for the larger stress, there is a jump after $t\approx{10^2}$ indicating largescale breaking of cages and then proceeds towards diffusion. Thus, the arrested state obtained for $\sigma_0/\sigma_d=1.547$ could be labelled as a marginally stable absorbed state, which gets destabilized and goes into eventual steady flow at the slightest increase of imposed stress. 

To analyse at a spatial scale the divergence of trajectories for the two applied shear stresses, we use two-dimensional local MSD maps, using a coarse-graining scale of $\ell=2.47$ which gives a mesh of $256 \times 256$; see the panels in Fig.\ref{fig3}(d), where the maps (A)-(E) in each row corresponds to five different times as marked in Fig.\ref{fig3}(a)-(c). The top row corresponds to the case of $\sigma_0/\sigma_d=1.547$, where there is no flow at the end and the system gets arrested, and the bottom row corresponds to the case of $\sigma_0/\sigma_d=1.557$ where we have a steady flow at the end. All displacements are measured relative to the initial state, which is the same for both cases. We clearly observe that even at the local scale, the MSD maps are very similar for (A)-(C), i.e. for $t= 0.06, 87.6, 167.7$. From (D) onward, i.e. $t=377.3$, the maps differ extensively. This indicates that for both the applied stresses the system evolves through similar trajectories in configuration space till the point of divergence. For the eventually flowing state, we clearly saw strong heterogeneity in the flow pattern in (D), with a vertical band of large mobility around a possible slip line. 

To characterize the local plasticity, we  compute maps of local vorticity $\omega = \nabla \times \tilde{\bf v}$, where $\tilde{\bf v}$ is the coarse-grained velocity field, using the same mesh size detailed above; these maps are shown in the panels in Fig.\ref{fig3}(e), for the both the stress magnitudies and for the same time points for which the MSD maps have been constructed. At earlier times (e.g. see (C)), there are only some sporadic vortices scattered here and there, in both cases. However, the main map of interest is (D), which for the case of  $\sigma_0/\sigma_d=1.547$, has no spatial signal, whereas for the case of $\sigma_0/\sigma_d=1.557$, we observe local vortices aligned along the percolating slip line discussed above. Thus, large scale flow for the larger stress proceeds via the formation of these spatially spanning structure consisting of local vortices. Other such vortical structure can also been seen elsewhere across the system in this case, indicating that plastic events have started to occur across the system as it transits to steady flow.

%{\em Investigating spatial point of stability/instability == compute the overall as well as spatial difference in trajectories.}

\subsection{Probing response at different coarse-graining scales}

Finally, we probe and characterise the response to stress at different scales, to obtain a multi-scale flavour of the ensuing plasticity. 

As a first step in this direction,  {we compute spatially coarse-grained self-overlap function, $Q_{\ell}(X,Y;t)$, measured at different times $t$. %where $X=\frac{n_{i}L}{n_{L}}$, $Y=\frac{n_{j}L}{n_{L}}$ with $i,j \in [1,n_{L}]$. 
For a given coarse-grained mesh-block of size $\ell$, centered at position $(X,Y)$, the self-overlap function is defined as
$Q_{\ell}(X,Y; t)=\sum_{i=1}^{N_{XY}} w(\delta{r}(t))$ where $\delta{r}(t)=|\Vec{r_{i}}(t)-\Vec{r_{i}}(0)|$ is the displacement from the initial state at $t=0$, $N_{XY}$ is the number of particles located in the mesh at $(X,Y)$ at time $t=0$, and the window-function, $w(\delta{r}(t))=1$ if $\delta{r}(t)\leq a$  or otherwise $w(\delta{r}(t))=0$. A threshold of $a\approx 0.2$  allows to pick up positional changes due to plasticity. %The definition of spatial self-overlap function ensures $Q_{\ell}(X,Y;0)=1$ for all $X,Y$ and $Q_{\ell}(X,Y;\tau_{F})\rightarrow b$ when an individual mesh-block fluidizes at $t=\tau_{F}$. 
Further, we define that a mesh has fluidized when $Q_{\ell}(X,Y; t) \leq 0.1$, which provides some kind of first passage time-scale for initiation 
of plastic activity.
%captures well the macroscopic fluidization, obtained at a scale $a=0.2$ for all $X,Y$. 
Via this, we obtain the timescale maps, $\tau_{F}(X,Y)$  where  $\tau_{F}$ is the fluidization timescale of each individual mesh-block.}. This exercise is done for different coarse-graining scales, viz. $\ell=19.76, 9.88, 4.94, 2.47$ which gives number of mesh-blocks to be, respectively, $n_{L}=32,64,128,256$.

For the F and NF trajectories discussed above, the corresponding timescale maps are shown in Fig.\ref{fig4}(a); see Appendix (Fig. A1 and Fig. A2) for examples of corresponding time evolution of $Q_{\ell}(X,Y;t)$.  In the case of  $\sigma_0/\sigma_d=1.547$, we observe an interesting feature, viz. that there exists a percolating backbone on the right hand side which does not yield over the observation time window $(t\approx3.39X10^{3})$. This implies that this region is resistant to yield at this applied stress, and therefore we infer that this rigid backbone leads to the final dynamical arrest as discussed in the context of Fig.\ref{fig3}. In contrast, this backbone fluidizes for $\sigma_0/\sigma_d=1.557$ within the observation window, which makes this a flowing state. Thus, it is this resistant backbone which makes the difference between arrest and flow. 

A pertinent question to ask is what triggers the breakdown of this backbone for a small increase in stress. Our analysis of the difference in activity (see Fig. A3 in Appendix) between the two trajectories seem to indicate that increased plastic activity, at early times, in the area around where the eventual slip-line emerges at later times (see sub-panel (D) in Fig.\ref{fig3}(d)), seems to contribute to the toppling of the region which is rigid for the marginally stable absorbed state at $\sigma_0/\sigma_d=1.547$. Whether this is the generic mechanism that triggers yielding of a marginally absorbed state needs to be
examined in more details, which will be explored in future work.

The diversity of local fluidization timescales is highlighted for smaller coarse-graining scales, e.g. in the case of the F trajectory, but gets smoothened out over larger scales, as expected. On the other hand, coarse-graining over larger $\ell$ brings into focus the robust backbone visible for the NF trajectory. Further, for $\sigma_0/\sigma_d=1.557$, when we visualize the vorticity map at $t=377.3$, using these different coarse-graining scales, complex spatio-temporal patterns are visible at nearly granular scales, which get smoothened out with increasing coarse-graining, with the slip line remaining the dominant feature; see Fig.\ref{fig4}(b). 

It is possible to obtain a measure of the dynamical heterogeneity, $\chi(t)$, from the instantaneous spatial fluctuations in $Q_{\ell}(X,Y; t)$. The data for  $\chi(t)$ is shown in in Fig.\ref{fig4}(c) (i)-(ii);  for the different coarse graining scales $\ell$ (labelled I-IV)  the data is shown in the top two panels for the F and NF trajectories discussed above. We observe that $\chi(t)$ is non-monotonic in both cases. For the trajectory where complete fluidization occurs, $\chi(t)$ decays to zero, while for the the case where the system gets arrested, $\chi(t)$ reaches a plateau capturing the extent of fluidization undergone till arrest. The peak scale of fluctuations is of course more for the least coarse-graining scale and for the choice of $a=0.2$, the peak in $\chi(t)$ occurs around $t\approx{30}$.  These fluctuations are indicators of the plastic events that lead the system towards some local minima (see Fig.\ref{fig3}, but not the large-scale plasticity that leads to the system to steady flow; this would correspond to larger $a$.)
The peak shifts to larger times for increasing choice of $a$, as can be seen in Fig.\ref{fig4}(c) (iii)-(iv) where
we show the fluctuations, labelled as $\chi_a(t)$, for different $a$ values.
%Finally, we compare the timescale map with the yield-stress map of the initial state. {\em to check -- whether the flow initiates at some soft spot and then spatial correlations emerge wherein initially rigid zones also get fluidized.}

\begin{figure*}[t]
	\centerline{\includegraphics[scale=0.25]{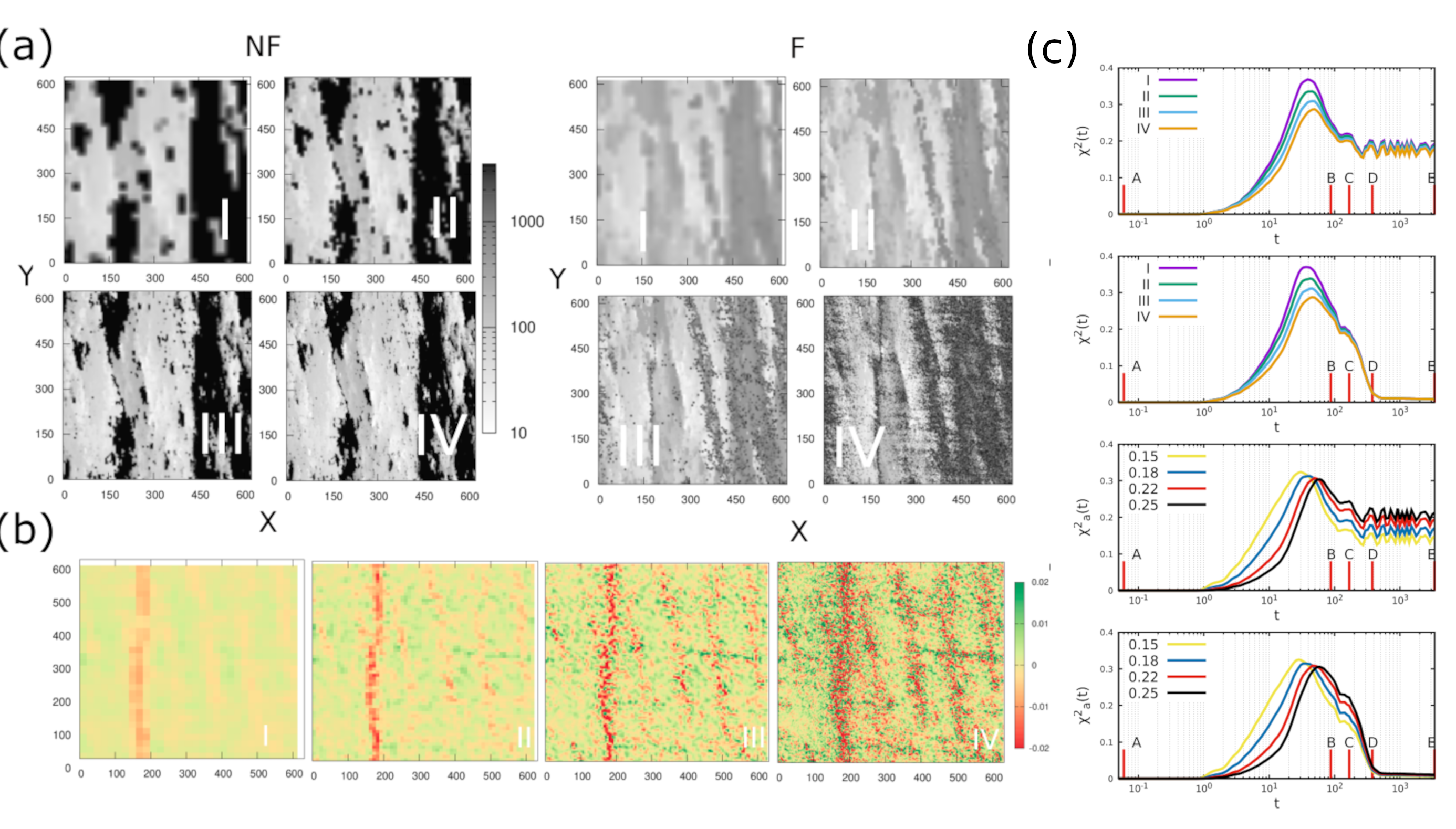}}
	\caption{{\bf Probing response at various lengthscales.}
		(a) Maps of local fluidization timescale $\tau_F$ computed at different coarse-grained scales $\ell=19.76$ (I), $9.88$ (II), $4.94$ (III), $2.47$ (IV), for (left) the non-fluidized (NF) state and (middle) fluidized (F) state, as indicated in Fig.\ref{fig3}. (b) Vorticity maps for the fluidized trajectory, measured at $t=377.3$ using the same $\ell$ scales as listed in (a). (c)  Spatial dynamical fluctuations, $\chi(t)$, computed via the overlap function $Q_{\ell}(X,Y; t)$, for the (i) non-fluidized and (ii) fluidized trajectories, using the $\ell$ scales labelled I-IV in (a). (iii)-(iv) Variation in fluctuations, $\chi_a(t)$, with choice of cutoff ($a$) in defining $Q_{\ell}(X,Y; t)$ for F and NF trajectories for different choices of $a$ as marked.}
	\label{fig4}
\end{figure*}

%\clearpage
\section{Summary and perspectives}

To summarize, we have studied the response of athermal amorphous states, which correspond to inherent structures of the model system, to applied shear stress. 

At first, we have studied the macroscopic response as well as corresponding fluctuations for different preparation histories of the inherent structures, which populate lower energy levels with increased annealing. The main observation, in this context, is that the yield threshold ($\sigma_y$) increases with decrease in the energy level of the initial undeformed state. For imposed stress values above the respective thresholds, the system reaches a long-time steady state flowing with a finite a shear-rate, which is of course independent of the history and compliant with the bulk flow curve. Below the threshold, the system reaches arrested states after some transient activity. Interestingly, the transient lifetime to eventual arrest or flow also depends upon history of preparation; more low lying states get arrested quickly if $\sigma_0 < \sigma_y$ and take longer to reach steady state when $\sigma_0 > \sigma_y$. The onset of fludization is also more {\em sudden} for such states. We show that the onset time for the large-scale plastic flow can be obtained by measuring the energy or strain fluctuations, across the ensemble of initial states:  maximal fluctuation mark the onset of large-scale plasticity and this increases as the imposed stress decreases towards $\sigma_y$.

Subsequently, we have done a spatio-temporal analysis of the response of the system for imposed stresses around the yield threshold, for the case of the lowest lying energy levels of the model that we could access via conventional annealing dynamics. For an imposed stress just below $\sigma_y$, we identified that the occurrence of percolating clusters resist yielding and this leads to dynamical arrest. These arrested states are marginally stable, since with a small increase in the applied stress, this rigid cluster breaks down due to the imposed stress and the system eventually fluidizes.  Our preliminary analysis indicates that the increases activity emanating from the soft spots leads to the local yielding and eventual flow. The onset of largescale flow occurs via spatio-temporal patterns that can be visualised in non-affine velocities as well as local vorticity, forming slip-plane like structures spanning the system. Furthermore, there is a wide variation of local yielding timescales during the onset of flow, which we characterise by quantifying the spatiotemporal fluctuations.

Finally, we note that our work can also be contextualised to the recent explorations of the relaxation dynamics within the energy landscapes of amorphous solids; e.g. see \cite{nishikawa2022relaxation, PhysRevLett.123.108001, vasisht2022residual, PhysRevLett.125.218001}. The distinction in our case is that the imposed stress alters the energy landscape and thus one studies the search dynamics in this altered space. However, similar to observations in some of these earlier studies, there is a power-law approach in the dynamics to arrest, as is manifested in the observed macroscopic shear-rate in the protocol that we study. We also note that a recent mean field study of fatigue behaviour in cyclic shear of amorphous systems reported yield rate curves which are very similar to shear-rate curves that we observe and similar bifurcation in the vicinity of an yield threshold \cite{PhysRevLett.128.198001}, suggesting some generic scenario of marginal stability around yield that needs to be explored further.

\section{Acknowledgements}

KM and PC acknowledge funding for this project via the CEFIPRA Project 5604-1. SD acknowledges support of the Department of Atomic Energy, Government of India, under project no RTI4001. We thank the HPC facility at IMSc Chennai for providing computational resources. We also thank P. Sollich, M. Ozawa for useful discussions. 

%\clearpage

\bibliography{biblo}

%merlin.mbs apsrev4-1.bst 2010-07-25 4.21a (PWD, AO, DPC) hacked
%Control: key (0)
%Control: author (8) initials jnrlst
%Control: editor formatted (1) identically to author
%Control: production of article title (-1) disabled
%Control: page (0) single
%Control: year (1) truncated
%Control: production of eprint (0) enabled
\providecommand{\noopsort}[1]{}\providecommand{\singleletter}[1]{#1}%
\begin{thebibliography}{47}%
\makeatletter
\providecommand \@ifxundefined [1]{%
 \@ifx{#1\undefined}
}%
\providecommand \@ifnum [1]{%
 \ifnum #1\expandafter \@firstoftwo
 \else \expandafter \@secondoftwo
 \fi
}%
\providecommand \@ifx [1]{%
 \ifx #1\expandafter \@firstoftwo
 \else \expandafter \@secondoftwo
 \fi
}%
\providecommand \natexlab [1]{#1}%
\providecommand \enquote  [1]{``#1''}%
\providecommand \bibnamefont  [1]{#1}%
\providecommand \bibfnamefont [1]{#1}%
\providecommand \citenamefont [1]{#1}%
\providecommand \href@noop [0]{\@secondoftwo}%
\providecommand \href [0]{\begingroup \@sanitize@url \@href}%
\providecommand \@href[1]{\@@startlink{#1}\@@href}%
\providecommand \@@href[1]{\endgroup#1\@@endlink}%
\providecommand \@sanitize@url [0]{\catcode `\\12\catcode `\$12\catcode
  `\&12\catcode `\#12\catcode `\^12\catcode `\_12\catcode `\%12\relax}%
\providecommand \@@startlink[1]{}%
\providecommand \@@endlink[0]{}%
\providecommand \url  [0]{\begingroup\@sanitize@url \@url }%
\providecommand \@url [1]{\endgroup\@href {#1}{\urlprefix }}%
\providecommand \urlprefix  [0]{URL }%
\providecommand \Eprint [0]{\href }%
\providecommand \doibase [0]{http://dx.doi.org/}%
\providecommand \selectlanguage [0]{\@gobble}%
\providecommand \bibinfo  [0]{\@secondoftwo}%
\providecommand \bibfield  [0]{\@secondoftwo}%
\providecommand \translation [1]{[#1]}%
\providecommand \BibitemOpen [0]{}%
\providecommand \bibitemStop [0]{}%
\providecommand \bibitemNoStop [0]{.\EOS\space}%
\providecommand \EOS [0]{\spacefactor3000\relax}%
\providecommand \BibitemShut  [1]{\csname bibitem#1\endcsname}%
\let\auto@bib@innerbib\@empty
%</preamble>
\bibitem [{\citenamefont {Schuh}\ \emph {et~al.}(2007)\citenamefont {Schuh},
  \citenamefont {Hufnagel},\ and\ \citenamefont
  {Ramamurty}}]{schuh2007mechanical}%
  \BibitemOpen
  \bibfield  {author} {\bibinfo {author} {\bibfnamefont {C.~A.}\ \bibnamefont
  {Schuh}}, \bibinfo {author} {\bibfnamefont {T.~C.}\ \bibnamefont {Hufnagel}},
  \ and\ \bibinfo {author} {\bibfnamefont {U.}~\bibnamefont {Ramamurty}},\
  }\href@noop {} {\bibfield  {journal} {\bibinfo  {journal} {Acta Materialia}\
  }\textbf {\bibinfo {volume} {55}},\ \bibinfo {pages} {4067} (\bibinfo {year}
  {2007})}\BibitemShut {NoStop}%
\bibitem [{\citenamefont {Rodney}\ \emph {et~al.}(2011)\citenamefont {Rodney},
  \citenamefont {Tanguy},\ and\ \citenamefont
  {Vandembroucq}}]{rodney2011modeling}%
  \BibitemOpen
  \bibfield  {author} {\bibinfo {author} {\bibfnamefont {D.}~\bibnamefont
  {Rodney}}, \bibinfo {author} {\bibfnamefont {A.}~\bibnamefont {Tanguy}}, \
  and\ \bibinfo {author} {\bibfnamefont {D.}~\bibnamefont {Vandembroucq}},\
  }\href@noop {} {\bibfield  {journal} {\bibinfo  {journal} {Modelling and
  Simulation in Materials Science and Engineering}\ }\textbf {\bibinfo {volume}
  {19}},\ \bibinfo {pages} {083001} (\bibinfo {year} {2011})}\BibitemShut
  {NoStop}%
\bibitem [{\citenamefont {Bonn}\ \emph {et~al.}(2017)\citenamefont {Bonn},
  \citenamefont {Denn}, \citenamefont {Berthier}, \citenamefont {Divoux},\ and\
  \citenamefont {Manneville}}]{bonn2017yield}%
  \BibitemOpen
  \bibfield  {author} {\bibinfo {author} {\bibfnamefont {D.}~\bibnamefont
  {Bonn}}, \bibinfo {author} {\bibfnamefont {M.~M.}\ \bibnamefont {Denn}},
  \bibinfo {author} {\bibfnamefont {L.}~\bibnamefont {Berthier}}, \bibinfo
  {author} {\bibfnamefont {T.}~\bibnamefont {Divoux}}, \ and\ \bibinfo {author}
  {\bibfnamefont {S.}~\bibnamefont {Manneville}},\ }\href@noop {} {\bibfield
  {journal} {\bibinfo  {journal} {Reviews of Modern Physics}\ }\textbf
  {\bibinfo {volume} {89}},\ \bibinfo {pages} {035005} (\bibinfo {year}
  {2017})}\BibitemShut {NoStop}%
\bibitem [{\citenamefont {Nicolas}\ \emph {et~al.}(2018)\citenamefont
  {Nicolas}, \citenamefont {Ferrero}, \citenamefont {Martens},\ and\
  \citenamefont {Barrat}}]{nicolas2018deformation}%
  \BibitemOpen
  \bibfield  {author} {\bibinfo {author} {\bibfnamefont {A.}~\bibnamefont
  {Nicolas}}, \bibinfo {author} {\bibfnamefont {E.~E.}\ \bibnamefont
  {Ferrero}}, \bibinfo {author} {\bibfnamefont {K.}~\bibnamefont {Martens}}, \
  and\ \bibinfo {author} {\bibfnamefont {J.-L.}\ \bibnamefont {Barrat}},\
  }\href@noop {} {\bibfield  {journal} {\bibinfo  {journal} {Reviews of Modern
  Physics}\ }\textbf {\bibinfo {volume} {90}},\ \bibinfo {pages} {045006}
  (\bibinfo {year} {2018})}\BibitemShut {NoStop}%
\bibitem [{\citenamefont {Fielding}(2014)}]{fielding2014shear}%
  \BibitemOpen
  \bibfield  {author} {\bibinfo {author} {\bibfnamefont {S.~M.}\ \bibnamefont
  {Fielding}},\ }\href@noop {} {\bibfield  {journal} {\bibinfo  {journal}
  {Reports on Progress in Physics}\ }\textbf {\bibinfo {volume} {77}},\
  \bibinfo {pages} {102601} (\bibinfo {year} {2014})}\BibitemShut {NoStop}%
\bibitem [{\citenamefont {Ozawa}\ \emph {et~al.}(2018)\citenamefont {Ozawa},
  \citenamefont {Berthier}, \citenamefont {Biroli}, \citenamefont {Rosso},\
  and\ \citenamefont {Tarjus}}]{ozawa2018random}%
  \BibitemOpen
  \bibfield  {author} {\bibinfo {author} {\bibfnamefont {M.}~\bibnamefont
  {Ozawa}}, \bibinfo {author} {\bibfnamefont {L.}~\bibnamefont {Berthier}},
  \bibinfo {author} {\bibfnamefont {G.}~\bibnamefont {Biroli}}, \bibinfo
  {author} {\bibfnamefont {A.}~\bibnamefont {Rosso}}, \ and\ \bibinfo {author}
  {\bibfnamefont {G.}~\bibnamefont {Tarjus}},\ }\href@noop {} {\bibfield
  {journal} {\bibinfo  {journal} {Proceedings of the National Academy of
  Sciences}\ }\textbf {\bibinfo {volume} {115}},\ \bibinfo {pages} {6656}
  (\bibinfo {year} {2018})}\BibitemShut {NoStop}%
\bibitem [{\citenamefont {Popovi{\'c}}\ \emph {et~al.}(2018)\citenamefont
  {Popovi{\'c}}, \citenamefont {de~Geus},\ and\ \citenamefont
  {Wyart}}]{popovic2018elastoplastic}%
  \BibitemOpen
  \bibfield  {author} {\bibinfo {author} {\bibfnamefont {M.}~\bibnamefont
  {Popovi{\'c}}}, \bibinfo {author} {\bibfnamefont {T.~W.}\ \bibnamefont
  {de~Geus}}, \ and\ \bibinfo {author} {\bibfnamefont {M.}~\bibnamefont
  {Wyart}},\ }\href@noop {} {\bibfield  {journal} {\bibinfo  {journal}
  {Physical Review E}\ }\textbf {\bibinfo {volume} {98}},\ \bibinfo {pages}
  {040901} (\bibinfo {year} {2018})}\BibitemShut {NoStop}%
\bibitem [{\citenamefont {Barlow}\ \emph {et~al.}(2020)\citenamefont {Barlow},
  \citenamefont {Cochran},\ and\ \citenamefont {Fielding}}]{barlow2020ductile}%
  \BibitemOpen
  \bibfield  {author} {\bibinfo {author} {\bibfnamefont {H.~J.}\ \bibnamefont
  {Barlow}}, \bibinfo {author} {\bibfnamefont {J.~O.}\ \bibnamefont {Cochran}},
  \ and\ \bibinfo {author} {\bibfnamefont {S.~M.}\ \bibnamefont {Fielding}},\
  }\href@noop {} {\bibfield  {journal} {\bibinfo  {journal} {Physical Review
  Letters}\ }\textbf {\bibinfo {volume} {125}},\ \bibinfo {pages} {168003}
  (\bibinfo {year} {2020})}\BibitemShut {NoStop}%
\bibitem [{\citenamefont {Richard}\ \emph
  {et~al.}(2021{\natexlab{a}})\citenamefont {Richard}, \citenamefont
  {Rainone},\ and\ \citenamefont {Lerner}}]{richardfinite}%
  \BibitemOpen
  \bibfield  {author} {\bibinfo {author} {\bibfnamefont {D.}~\bibnamefont
  {Richard}}, \bibinfo {author} {\bibfnamefont {C.}~\bibnamefont {Rainone}}, \
  and\ \bibinfo {author} {\bibfnamefont {E.}~\bibnamefont {Lerner}},\ }\href
  {\doibase 10.1063/5.0053303} {\bibfield  {journal} {\bibinfo  {journal} {The
  Journal of Chemical Physics}\ }\textbf {\bibinfo {volume} {155}},\ \bibinfo
  {pages} {056101} (\bibinfo {year} {2021}{\natexlab{a}})},\ \Eprint
  {http://arxiv.org/abs/https://doi.org/10.1063/5.0053303}
  {https://doi.org/10.1063/5.0053303} \BibitemShut {NoStop}%
\bibitem [{\citenamefont {Richard}\ \emph
  {et~al.}(2021{\natexlab{b}})\citenamefont {Richard}, \citenamefont {Lerner},\
  and\ \citenamefont {Bouchbinder}}]{richardbrittle2021}%
  \BibitemOpen
  \bibfield  {author} {\bibinfo {author} {\bibfnamefont {D.}~\bibnamefont
  {Richard}}, \bibinfo {author} {\bibfnamefont {E.}~\bibnamefont {Lerner}}, \
  and\ \bibinfo {author} {\bibfnamefont {E.}~\bibnamefont {Bouchbinder}},\
  }\href@noop {} {\bibfield  {journal} {\bibinfo  {journal} {MRS Bulletin}\
  }\textbf {\bibinfo {volume} {46}},\ \bibinfo {pages} {902} (\bibinfo {year}
  {2021}{\natexlab{b}})}\BibitemShut {NoStop}%
\bibitem [{\citenamefont {Rossi}\ \emph {et~al.}(2022)\citenamefont {Rossi},
  \citenamefont {Biroli}, \citenamefont {Ozawa}, \citenamefont {Tarjus},\ and\
  \citenamefont {Zamponi}}]{rossi2022finite}%
  \BibitemOpen
  \bibfield  {author} {\bibinfo {author} {\bibfnamefont {S.}~\bibnamefont
  {Rossi}}, \bibinfo {author} {\bibfnamefont {G.}~\bibnamefont {Biroli}},
  \bibinfo {author} {\bibfnamefont {M.}~\bibnamefont {Ozawa}}, \bibinfo
  {author} {\bibfnamefont {G.}~\bibnamefont {Tarjus}}, \ and\ \bibinfo {author}
  {\bibfnamefont {F.}~\bibnamefont {Zamponi}},\ }\href@noop {} {\bibfield
  {journal} {\bibinfo  {journal} {Physical Review Letters}\ }\textbf {\bibinfo
  {volume} {129}},\ \bibinfo {pages} {228002} (\bibinfo {year}
  {2022})}\BibitemShut {NoStop}%
\bibitem [{\citenamefont {Pollard}\ and\ \citenamefont
  {Fielding}(2022)}]{pollard2022yielding}%
  \BibitemOpen
  \bibfield  {author} {\bibinfo {author} {\bibfnamefont {J.}~\bibnamefont
  {Pollard}}\ and\ \bibinfo {author} {\bibfnamefont {S.~M.}\ \bibnamefont
  {Fielding}},\ }\href@noop {} {\bibfield  {journal} {\bibinfo  {journal}
  {Physical Review Research}\ }\textbf {\bibinfo {volume} {4}},\ \bibinfo
  {pages} {043037} (\bibinfo {year} {2022})}\BibitemShut {NoStop}%
\bibitem [{\citenamefont {Betten}(2008)}]{betten2008creep}%
  \BibitemOpen
  \bibfield  {author} {\bibinfo {author} {\bibfnamefont {J.}~\bibnamefont
  {Betten}},\ }\href@noop {} {\emph {\bibinfo {title} {Creep mechanics}}}\
  (\bibinfo  {publisher} {Springer Science \& Business Media},\ \bibinfo {year}
  {2008})\BibitemShut {NoStop}%
\bibitem [{\citenamefont {Bauer}\ \emph {et~al.}(2006)\citenamefont {Bauer},
  \citenamefont {Oberdisse},\ and\ \citenamefont
  {Ramos}}]{bauer2006collective}%
  \BibitemOpen
  \bibfield  {author} {\bibinfo {author} {\bibfnamefont {T.}~\bibnamefont
  {Bauer}}, \bibinfo {author} {\bibfnamefont {J.}~\bibnamefont {Oberdisse}}, \
  and\ \bibinfo {author} {\bibfnamefont {L.}~\bibnamefont {Ramos}},\
  }\href@noop {} {\bibfield  {journal} {\bibinfo  {journal} {Physical review
  letters}\ }\textbf {\bibinfo {volume} {97}},\ \bibinfo {pages} {258303}
  (\bibinfo {year} {2006})}\BibitemShut {NoStop}%
\bibitem [{\citenamefont {Divoux}\ \emph
  {et~al.}(2011{\natexlab{a}})\citenamefont {Divoux}, \citenamefont
  {Barentin},\ and\ \citenamefont {Manneville}}]{Divoux2011a}%
  \BibitemOpen
  \bibfield  {author} {\bibinfo {author} {\bibfnamefont {T.}~\bibnamefont
  {Divoux}}, \bibinfo {author} {\bibfnamefont {C.}~\bibnamefont {Barentin}}, \
  and\ \bibinfo {author} {\bibfnamefont {S.}~\bibnamefont {Manneville}},\
  }\href@noop {} {\bibfield  {journal} {\bibinfo  {journal} {Soft Matter}\
  }\textbf {\bibinfo {volume} {7}},\ \bibinfo {pages} {9335} (\bibinfo {year}
  {2011}{\natexlab{a}})}\BibitemShut {NoStop}%
\bibitem [{\citenamefont {Leocmach}\ \emph {et~al.}(2014)\citenamefont
  {Leocmach}, \citenamefont {Perge}, \citenamefont {Divoux},\ and\
  \citenamefont {Manneville}}]{Leocmach2014}%
  \BibitemOpen
  \bibfield  {author} {\bibinfo {author} {\bibfnamefont {M.}~\bibnamefont
  {Leocmach}}, \bibinfo {author} {\bibfnamefont {C.}~\bibnamefont {Perge}},
  \bibinfo {author} {\bibfnamefont {T.}~\bibnamefont {Divoux}}, \ and\ \bibinfo
  {author} {\bibfnamefont {S.}~\bibnamefont {Manneville}},\ }\href {\doibase
  10.1103/PhysRevLett.113.038303} {\bibfield  {journal} {\bibinfo  {journal}
  {Physical Review Letters}\ }\textbf {\bibinfo {volume} {113}},\ \bibinfo
  {pages} {1} (\bibinfo {year} {2014})},\ \Eprint
  {http://arxiv.org/abs/1401.8234} {arXiv:1401.8234} \BibitemShut {NoStop}%
\bibitem [{\citenamefont {Ballesta}\ and\ \citenamefont
  {Petekidis}(2016)}]{ballesta2016creep}%
  \BibitemOpen
  \bibfield  {author} {\bibinfo {author} {\bibfnamefont {P.}~\bibnamefont
  {Ballesta}}\ and\ \bibinfo {author} {\bibfnamefont {G.}~\bibnamefont
  {Petekidis}},\ }\href@noop {} {\bibfield  {journal} {\bibinfo  {journal}
  {Physical Review E}\ }\textbf {\bibinfo {volume} {93}},\ \bibinfo {pages}
  {042613} (\bibinfo {year} {2016})}\BibitemShut {NoStop}%
\bibitem [{\citenamefont {Chaudhuri}\ and\ \citenamefont
  {Horbach}(2013)}]{chaudhuri2013onset}%
  \BibitemOpen
  \bibfield  {author} {\bibinfo {author} {\bibfnamefont {P.}~\bibnamefont
  {Chaudhuri}}\ and\ \bibinfo {author} {\bibfnamefont {J.}~\bibnamefont
  {Horbach}},\ }\href@noop {} {\bibfield  {journal} {\bibinfo  {journal}
  {Physical Review E}\ }\textbf {\bibinfo {volume} {88}},\ \bibinfo {pages}
  {040301} (\bibinfo {year} {2013})}\BibitemShut {NoStop}%
\bibitem [{\citenamefont {Sentjabrskaja}\ \emph {et~al.}(2015)\citenamefont
  {Sentjabrskaja}, \citenamefont {Chaudhuri}, \citenamefont {Hermes},
  \citenamefont {Poon}, \citenamefont {Horbach}, \citenamefont {Egelhaaf},\
  and\ \citenamefont {Laurati}}]{sentjabrskaja2015creep}%
  \BibitemOpen
  \bibfield  {author} {\bibinfo {author} {\bibfnamefont {T.}~\bibnamefont
  {Sentjabrskaja}}, \bibinfo {author} {\bibfnamefont {P.}~\bibnamefont
  {Chaudhuri}}, \bibinfo {author} {\bibfnamefont {M.}~\bibnamefont {Hermes}},
  \bibinfo {author} {\bibfnamefont {W.}~\bibnamefont {Poon}}, \bibinfo {author}
  {\bibfnamefont {J.}~\bibnamefont {Horbach}}, \bibinfo {author} {\bibfnamefont
  {S.}~\bibnamefont {Egelhaaf}}, \ and\ \bibinfo {author} {\bibfnamefont
  {M.}~\bibnamefont {Laurati}},\ }\href@noop {} {\bibfield  {journal} {\bibinfo
   {journal} {Scientific reports}\ }\textbf {\bibinfo {volume} {5}},\ \bibinfo
  {pages} {11884} (\bibinfo {year} {2015})}\BibitemShut {NoStop}%
\bibitem [{\citenamefont {Landrum}\ \emph {et~al.}(2016)\citenamefont
  {Landrum}, \citenamefont {Russel},\ and\ \citenamefont
  {Zia}}]{landrum2016delayed}%
  \BibitemOpen
  \bibfield  {author} {\bibinfo {author} {\bibfnamefont {B.~J.}\ \bibnamefont
  {Landrum}}, \bibinfo {author} {\bibfnamefont {W.~B.}\ \bibnamefont {Russel}},
  \ and\ \bibinfo {author} {\bibfnamefont {R.~N.}\ \bibnamefont {Zia}},\
  }\href@noop {} {\bibfield  {journal} {\bibinfo  {journal} {Journal of
  Rheology}\ }\textbf {\bibinfo {volume} {60}},\ \bibinfo {pages} {783}
  (\bibinfo {year} {2016})}\BibitemShut {NoStop}%
\bibitem [{\citenamefont {Cipelletti}\ \emph {et~al.}(2020)\citenamefont
  {Cipelletti}, \citenamefont {Martens},\ and\ \citenamefont
  {Ramos}}]{cipelletti2020microscopic}%
  \BibitemOpen
  \bibfield  {author} {\bibinfo {author} {\bibfnamefont {L.}~\bibnamefont
  {Cipelletti}}, \bibinfo {author} {\bibfnamefont {K.}~\bibnamefont {Martens}},
  \ and\ \bibinfo {author} {\bibfnamefont {L.}~\bibnamefont {Ramos}},\
  }\href@noop {} {\bibfield  {journal} {\bibinfo  {journal} {Soft matter}\
  }\textbf {\bibinfo {volume} {16}},\ \bibinfo {pages} {82} (\bibinfo {year}
  {2020})}\BibitemShut {NoStop}%
\bibitem [{\citenamefont {Merabia}\ and\ \citenamefont
  {Detcheverry}(2016)}]{merabia2016thermally}%
  \BibitemOpen
  \bibfield  {author} {\bibinfo {author} {\bibfnamefont {S.}~\bibnamefont
  {Merabia}}\ and\ \bibinfo {author} {\bibfnamefont {F.}~\bibnamefont
  {Detcheverry}},\ }\href@noop {} {\bibfield  {journal} {\bibinfo  {journal}
  {EPL (Europhysics Letters)}\ }\textbf {\bibinfo {volume} {116}},\ \bibinfo
  {pages} {46003} (\bibinfo {year} {2016})}\BibitemShut {NoStop}%
\bibitem [{\citenamefont {Joshi}\ and\ \citenamefont
  {Petekidis}(2018)}]{joshi2018yield}%
  \BibitemOpen
  \bibfield  {author} {\bibinfo {author} {\bibfnamefont {Y.~M.}\ \bibnamefont
  {Joshi}}\ and\ \bibinfo {author} {\bibfnamefont {G.}~\bibnamefont
  {Petekidis}},\ }\href@noop {} {\bibfield  {journal} {\bibinfo  {journal}
  {Rheologica Acta}\ }\textbf {\bibinfo {volume} {57}},\ \bibinfo {pages} {521}
  (\bibinfo {year} {2018})}\BibitemShut {NoStop}%
\bibitem [{\citenamefont {Cabriolu}\ \emph {et~al.}(2019)\citenamefont
  {Cabriolu}, \citenamefont {Horbach}, \citenamefont {Chaudhuri},\ and\
  \citenamefont {Martens}}]{cabriolu2019precursors}%
  \BibitemOpen
  \bibfield  {author} {\bibinfo {author} {\bibfnamefont {R.}~\bibnamefont
  {Cabriolu}}, \bibinfo {author} {\bibfnamefont {J.}~\bibnamefont {Horbach}},
  \bibinfo {author} {\bibfnamefont {P.}~\bibnamefont {Chaudhuri}}, \ and\
  \bibinfo {author} {\bibfnamefont {K.}~\bibnamefont {Martens}},\ }\href@noop
  {} {\bibfield  {journal} {\bibinfo  {journal} {Soft matter}\ }\textbf
  {\bibinfo {volume} {15}},\ \bibinfo {pages} {415} (\bibinfo {year}
  {2019})}\BibitemShut {NoStop}%
\bibitem [{\citenamefont {Dijksman}\ and\ \citenamefont
  {Mullin}(2022)}]{dijksman2022creep}%
  \BibitemOpen
  \bibfield  {author} {\bibinfo {author} {\bibfnamefont {J.~A.}\ \bibnamefont
  {Dijksman}}\ and\ \bibinfo {author} {\bibfnamefont {T.}~\bibnamefont
  {Mullin}},\ }\href@noop {} {\bibfield  {journal} {\bibinfo  {journal}
  {Physical Review Letters}\ }\textbf {\bibinfo {volume} {128}},\ \bibinfo
  {pages} {238002} (\bibinfo {year} {2022})}\BibitemShut {NoStop}%
\bibitem [{\citenamefont {Bhowmik}\ \emph {et~al.}(2022)\citenamefont
  {Bhowmik}, \citenamefont {Hentschel},\ and\ \citenamefont
  {Procaccia}}]{bhowmik2022creep}%
  \BibitemOpen
  \bibfield  {author} {\bibinfo {author} {\bibfnamefont {B.~P.}\ \bibnamefont
  {Bhowmik}}, \bibinfo {author} {\bibfnamefont {H.}~\bibnamefont {Hentschel}},
  \ and\ \bibinfo {author} {\bibfnamefont {I.}~\bibnamefont {Procaccia}},\
  }\href@noop {} {\bibfield  {journal} {\bibinfo  {journal} {Physical Review
  E}\ }\textbf {\bibinfo {volume} {106}},\ \bibinfo {pages} {034906} (\bibinfo
  {year} {2022})}\BibitemShut {NoStop}%
\bibitem [{\citenamefont {Siebenb{\"u}rger}\ \emph {et~al.}(2012)\citenamefont
  {Siebenb{\"u}rger}, \citenamefont {Ballauff},\ and\ \citenamefont
  {Voigtmann}}]{siebenburger2012creep}%
  \BibitemOpen
  \bibfield  {author} {\bibinfo {author} {\bibfnamefont {M.}~\bibnamefont
  {Siebenb{\"u}rger}}, \bibinfo {author} {\bibfnamefont {M.}~\bibnamefont
  {Ballauff}}, \ and\ \bibinfo {author} {\bibfnamefont {T.}~\bibnamefont
  {Voigtmann}},\ }\href@noop {} {\bibfield  {journal} {\bibinfo  {journal}
  {Physical review letters}\ }\textbf {\bibinfo {volume} {108}},\ \bibinfo
  {pages} {255701} (\bibinfo {year} {2012})}\BibitemShut {NoStop}%
\bibitem [{\citenamefont {Paredes}\ \emph {et~al.}(2013)\citenamefont
  {Paredes}, \citenamefont {Michels},\ and\ \citenamefont
  {Bonn}}]{paredes2013rheology}%
  \BibitemOpen
  \bibfield  {author} {\bibinfo {author} {\bibfnamefont {J.}~\bibnamefont
  {Paredes}}, \bibinfo {author} {\bibfnamefont {M.~A.}\ \bibnamefont
  {Michels}}, \ and\ \bibinfo {author} {\bibfnamefont {D.}~\bibnamefont
  {Bonn}},\ }\href@noop {} {\bibfield  {journal} {\bibinfo  {journal} {Physical
  review letters}\ }\textbf {\bibinfo {volume} {111}},\ \bibinfo {pages}
  {015701} (\bibinfo {year} {2013})}\BibitemShut {NoStop}%
\bibitem [{\citenamefont {Liu}\ \emph {et~al.}(2018{\natexlab{a}})\citenamefont
  {Liu}, \citenamefont {Martens},\ and\ \citenamefont
  {Barrat}}]{PhysRevLett.120.028004}%
  \BibitemOpen
  \bibfield  {author} {\bibinfo {author} {\bibfnamefont {C.}~\bibnamefont
  {Liu}}, \bibinfo {author} {\bibfnamefont {K.}~\bibnamefont {Martens}}, \ and\
  \bibinfo {author} {\bibfnamefont {J.-L.}\ \bibnamefont {Barrat}},\ }\href
  {\doibase 10.1103/PhysRevLett.120.028004} {\bibfield  {journal} {\bibinfo
  {journal} {Phys. Rev. Lett.}\ }\textbf {\bibinfo {volume} {120}},\ \bibinfo
  {pages} {028004} (\bibinfo {year} {2018}{\natexlab{a}})}\BibitemShut
  {NoStop}%
\bibitem [{\citenamefont {Liu}\ \emph {et~al.}(2018{\natexlab{b}})\citenamefont
  {Liu}, \citenamefont {Ferrero}, \citenamefont {Martens},\ and\ \citenamefont
  {Barrat}}]{liu2018creep}%
  \BibitemOpen
  \bibfield  {author} {\bibinfo {author} {\bibfnamefont {C.}~\bibnamefont
  {Liu}}, \bibinfo {author} {\bibfnamefont {E.~E.}\ \bibnamefont {Ferrero}},
  \bibinfo {author} {\bibfnamefont {K.}~\bibnamefont {Martens}}, \ and\
  \bibinfo {author} {\bibfnamefont {J.-L.}\ \bibnamefont {Barrat}},\
  }\href@noop {} {\bibfield  {journal} {\bibinfo  {journal} {Soft matter}\
  }\textbf {\bibinfo {volume} {14}},\ \bibinfo {pages} {8306} (\bibinfo {year}
  {2018}{\natexlab{b}})}\BibitemShut {NoStop}%
\bibitem [{\citenamefont {Liu}\ \emph {et~al.}(2021)\citenamefont {Liu},
  \citenamefont {Dutta}, \citenamefont {Chaudhuri},\ and\ \citenamefont
  {Martens}}]{liu2021elastoplastic}%
  \BibitemOpen
  \bibfield  {author} {\bibinfo {author} {\bibfnamefont {C.}~\bibnamefont
  {Liu}}, \bibinfo {author} {\bibfnamefont {S.}~\bibnamefont {Dutta}}, \bibinfo
  {author} {\bibfnamefont {P.}~\bibnamefont {Chaudhuri}}, \ and\ \bibinfo
  {author} {\bibfnamefont {K.}~\bibnamefont {Martens}},\ }\href@noop {}
  {\bibfield  {journal} {\bibinfo  {journal} {Physical Review Letters}\
  }\textbf {\bibinfo {volume} {126}},\ \bibinfo {pages} {138005} (\bibinfo
  {year} {2021})}\BibitemShut {NoStop}%
\bibitem [{\citenamefont {Popovi{\'c}}\ \emph {et~al.}(2022)\citenamefont
  {Popovi{\'c}}, \citenamefont {de~Geus}, \citenamefont {Ji}, \citenamefont
  {Rosso},\ and\ \citenamefont {Wyart}}]{popovic2022scaling}%
  \BibitemOpen
  \bibfield  {author} {\bibinfo {author} {\bibfnamefont {M.}~\bibnamefont
  {Popovi{\'c}}}, \bibinfo {author} {\bibfnamefont {T.~W.}\ \bibnamefont
  {de~Geus}}, \bibinfo {author} {\bibfnamefont {W.}~\bibnamefont {Ji}},
  \bibinfo {author} {\bibfnamefont {A.}~\bibnamefont {Rosso}}, \ and\ \bibinfo
  {author} {\bibfnamefont {M.}~\bibnamefont {Wyart}},\ }\href@noop {}
  {\bibfield  {journal} {\bibinfo  {journal} {Physical Review Letters}\
  }\textbf {\bibinfo {volume} {129}},\ \bibinfo {pages} {208001} (\bibinfo
  {year} {2022})}\BibitemShut {NoStop}%
\bibitem [{\citenamefont {Lan{\c{c}}on}\ and\ \citenamefont
  {Billard}(1988)}]{lanccon1988two}%
  \BibitemOpen
  \bibfield  {author} {\bibinfo {author} {\bibfnamefont {F.}~\bibnamefont
  {Lan{\c{c}}on}}\ and\ \bibinfo {author} {\bibfnamefont {L.}~\bibnamefont
  {Billard}},\ }\href@noop {} {\bibfield  {journal} {\bibinfo  {journal}
  {Journal de Physique}\ }\textbf {\bibinfo {volume} {49}},\ \bibinfo {pages}
  {249} (\bibinfo {year} {1988})}\BibitemShut {NoStop}%
\bibitem [{\citenamefont {Lan{\c{c}}on}\ \emph {et~al.}(1986)\citenamefont
  {Lan{\c{c}}on}, \citenamefont {Billard},\ and\ \citenamefont
  {Chaudhari}}]{lanccon1986thermodynamical}%
  \BibitemOpen
  \bibfield  {author} {\bibinfo {author} {\bibfnamefont {F.}~\bibnamefont
  {Lan{\c{c}}on}}, \bibinfo {author} {\bibfnamefont {L.}~\bibnamefont
  {Billard}}, \ and\ \bibinfo {author} {\bibfnamefont {P.}~\bibnamefont
  {Chaudhari}},\ }\href@noop {} {\bibfield  {journal} {\bibinfo  {journal} {EPL
  (Europhysics Letters)}\ }\textbf {\bibinfo {volume} {2}},\ \bibinfo {pages}
  {625} (\bibinfo {year} {1986})}\BibitemShut {NoStop}%
\bibitem [{\citenamefont {Falk}\ and\ \citenamefont
  {Langer}(1998)}]{falk1998dynamics}%
  \BibitemOpen
  \bibfield  {author} {\bibinfo {author} {\bibfnamefont {M.~L.}\ \bibnamefont
  {Falk}}\ and\ \bibinfo {author} {\bibfnamefont {J.~S.}\ \bibnamefont
  {Langer}},\ }\href@noop {} {\bibfield  {journal} {\bibinfo  {journal}
  {Physical Review E}\ }\textbf {\bibinfo {volume} {57}},\ \bibinfo {pages}
  {7192} (\bibinfo {year} {1998})}\BibitemShut {NoStop}%
\bibitem [{\citenamefont {Patinet}\ \emph {et~al.}(2016)\citenamefont
  {Patinet}, \citenamefont {Vandembroucq},\ and\ \citenamefont
  {Falk}}]{patinet2016connecting}%
  \BibitemOpen
  \bibfield  {author} {\bibinfo {author} {\bibfnamefont {S.}~\bibnamefont
  {Patinet}}, \bibinfo {author} {\bibfnamefont {D.}~\bibnamefont
  {Vandembroucq}}, \ and\ \bibinfo {author} {\bibfnamefont {M.~L.}\
  \bibnamefont {Falk}},\ }\href@noop {} {\bibfield  {journal} {\bibinfo
  {journal} {Physical review letters}\ }\textbf {\bibinfo {volume} {117}},\
  \bibinfo {pages} {045501} (\bibinfo {year} {2016})}\BibitemShut {NoStop}%
\bibitem [{\citenamefont {Barbot}\ \emph {et~al.}(2018)\citenamefont {Barbot},
  \citenamefont {Lerbinger}, \citenamefont {Hernandez-Garcia}, \citenamefont
  {Garc{\'\i}a-Garc{\'\i}a}, \citenamefont {Falk}, \citenamefont
  {Vandembroucq},\ and\ \citenamefont {Patinet}}]{barbot2018local}%
  \BibitemOpen
  \bibfield  {author} {\bibinfo {author} {\bibfnamefont {A.}~\bibnamefont
  {Barbot}}, \bibinfo {author} {\bibfnamefont {M.}~\bibnamefont {Lerbinger}},
  \bibinfo {author} {\bibfnamefont {A.}~\bibnamefont {Hernandez-Garcia}},
  \bibinfo {author} {\bibfnamefont {R.}~\bibnamefont
  {Garc{\'\i}a-Garc{\'\i}a}}, \bibinfo {author} {\bibfnamefont {M.~L.}\
  \bibnamefont {Falk}}, \bibinfo {author} {\bibfnamefont {D.}~\bibnamefont
  {Vandembroucq}}, \ and\ \bibinfo {author} {\bibfnamefont {S.}~\bibnamefont
  {Patinet}},\ }\href@noop {} {\bibfield  {journal} {\bibinfo  {journal}
  {Physical Review E}\ }\textbf {\bibinfo {volume} {97}},\ \bibinfo {pages}
  {033001} (\bibinfo {year} {2018})}\BibitemShut {NoStop}%
\bibitem [{\citenamefont {Patinet}\ \emph {et~al.}(2019)\citenamefont
  {Patinet}, \citenamefont {Barbot}, \citenamefont {Lerbinger}, \citenamefont
  {Vandembroucq},\ and\ \citenamefont {Lema{\^\i}tre}}]{patinet2019origin}%
  \BibitemOpen
  \bibfield  {author} {\bibinfo {author} {\bibfnamefont {S.}~\bibnamefont
  {Patinet}}, \bibinfo {author} {\bibfnamefont {A.}~\bibnamefont {Barbot}},
  \bibinfo {author} {\bibfnamefont {M.}~\bibnamefont {Lerbinger}}, \bibinfo
  {author} {\bibfnamefont {D.}~\bibnamefont {Vandembroucq}}, \ and\ \bibinfo
  {author} {\bibfnamefont {A.}~\bibnamefont {Lema{\^\i}tre}},\ }\href@noop {}
  {\bibfield  {journal} {\bibinfo  {journal} {arXiv preprint arXiv:1906.09818}\
  } (\bibinfo {year} {2019})}\BibitemShut {NoStop}%
\bibitem [{\citenamefont {Vezirov}\ \emph {et~al.}(2015)\citenamefont
  {Vezirov}, \citenamefont {Gerloff},\ and\ \citenamefont
  {Klapp}}]{vezirov2015manipulating}%
  \BibitemOpen
  \bibfield  {author} {\bibinfo {author} {\bibfnamefont {T.~A.}\ \bibnamefont
  {Vezirov}}, \bibinfo {author} {\bibfnamefont {S.}~\bibnamefont {Gerloff}}, \
  and\ \bibinfo {author} {\bibfnamefont {S.~H.}\ \bibnamefont {Klapp}},\
  }\href@noop {} {\bibfield  {journal} {\bibinfo  {journal} {Soft Matter}\
  }\textbf {\bibinfo {volume} {11}},\ \bibinfo {pages} {406} (\bibinfo {year}
  {2015})}\BibitemShut {NoStop}%
\bibitem [{\citenamefont {Plimpton}(1995)}]{plimpton1995fast}%
  \BibitemOpen
  \bibfield  {author} {\bibinfo {author} {\bibfnamefont {S.}~\bibnamefont
  {Plimpton}},\ }\href@noop {} {\bibfield  {journal} {\bibinfo  {journal}
  {Journal of computational physics}\ }\textbf {\bibinfo {volume} {117}},\
  \bibinfo {pages} {1} (\bibinfo {year} {1995})}\BibitemShut {NoStop}%
\bibitem [{\citenamefont {Divoux}\ \emph
  {et~al.}(2011{\natexlab{b}})\citenamefont {Divoux}, \citenamefont
  {Barentin},\ and\ \citenamefont {Manneville}}]{divoux2011stress}%
  \BibitemOpen
  \bibfield  {author} {\bibinfo {author} {\bibfnamefont {T.}~\bibnamefont
  {Divoux}}, \bibinfo {author} {\bibfnamefont {C.}~\bibnamefont {Barentin}}, \
  and\ \bibinfo {author} {\bibfnamefont {S.}~\bibnamefont {Manneville}},\
  }\href@noop {} {\bibfield  {journal} {\bibinfo  {journal} {Soft Matter}\
  }\textbf {\bibinfo {volume} {7}},\ \bibinfo {pages} {9335} (\bibinfo {year}
  {2011}{\natexlab{b}})}\BibitemShut {NoStop}%
\bibitem [{\citenamefont {Lamp}\ \emph {et~al.}(2022)\citenamefont {Lamp},
  \citenamefont {K{\"u}chler},\ and\ \citenamefont
  {Horbach}}]{lamp2022brittle}%
  \BibitemOpen
  \bibfield  {author} {\bibinfo {author} {\bibfnamefont {K.}~\bibnamefont
  {Lamp}}, \bibinfo {author} {\bibfnamefont {N.}~\bibnamefont {K{\"u}chler}}, \
  and\ \bibinfo {author} {\bibfnamefont {J.}~\bibnamefont {Horbach}},\
  }\href@noop {} {\bibfield  {journal} {\bibinfo  {journal} {The Journal of
  Chemical Physics}\ }\textbf {\bibinfo {volume} {157}},\ \bibinfo {pages}
  {034501} (\bibinfo {year} {2022})}\BibitemShut {NoStop}%
\bibitem [{\citenamefont {Nishikawa}\ \emph {et~al.}(2022)\citenamefont
  {Nishikawa}, \citenamefont {Ozawa}, \citenamefont {Ikeda}, \citenamefont
  {Chaudhuri},\ and\ \citenamefont {Berthier}}]{nishikawa2022relaxation}%
  \BibitemOpen
  \bibfield  {author} {\bibinfo {author} {\bibfnamefont {Y.}~\bibnamefont
  {Nishikawa}}, \bibinfo {author} {\bibfnamefont {M.}~\bibnamefont {Ozawa}},
  \bibinfo {author} {\bibfnamefont {A.}~\bibnamefont {Ikeda}}, \bibinfo
  {author} {\bibfnamefont {P.}~\bibnamefont {Chaudhuri}}, \ and\ \bibinfo
  {author} {\bibfnamefont {L.}~\bibnamefont {Berthier}},\ }\href@noop {}
  {\bibfield  {journal} {\bibinfo  {journal} {Physical Review X}\ }\textbf
  {\bibinfo {volume} {12}},\ \bibinfo {pages} {021001} (\bibinfo {year}
  {2022})}\BibitemShut {NoStop}%
\bibitem [{\citenamefont {Chacko}\ \emph {et~al.}(2019)\citenamefont {Chacko},
  \citenamefont {Sollich},\ and\ \citenamefont
  {Fielding}}]{PhysRevLett.123.108001}%
  \BibitemOpen
  \bibfield  {author} {\bibinfo {author} {\bibfnamefont {R.~N.}\ \bibnamefont
  {Chacko}}, \bibinfo {author} {\bibfnamefont {P.}~\bibnamefont {Sollich}}, \
  and\ \bibinfo {author} {\bibfnamefont {S.~M.}\ \bibnamefont {Fielding}},\
  }\href {\doibase 10.1103/PhysRevLett.123.108001} {\bibfield  {journal}
  {\bibinfo  {journal} {Phys. Rev. Lett.}\ }\textbf {\bibinfo {volume} {123}},\
  \bibinfo {pages} {108001} (\bibinfo {year} {2019})}\BibitemShut {NoStop}%
\bibitem [{\citenamefont {Vasisht}\ \emph {et~al.}(2022)\citenamefont
  {Vasisht}, \citenamefont {Chaudhuri},\ and\ \citenamefont
  {Martens}}]{vasisht2022residual}%
  \BibitemOpen
  \bibfield  {author} {\bibinfo {author} {\bibfnamefont {V.~V.}\ \bibnamefont
  {Vasisht}}, \bibinfo {author} {\bibfnamefont {P.}~\bibnamefont {Chaudhuri}},
  \ and\ \bibinfo {author} {\bibfnamefont {K.}~\bibnamefont {Martens}},\
  }\href@noop {} {\bibfield  {journal} {\bibinfo  {journal} {Soft Matter}\
  }\textbf {\bibinfo {volume} {18}},\ \bibinfo {pages} {6426} (\bibinfo {year}
  {2022})}\BibitemShut {NoStop}%
\bibitem [{\citenamefont {Mandal}\ and\ \citenamefont
  {Sollich}(2020)}]{PhysRevLett.125.218001}%
  \BibitemOpen
  \bibfield  {author} {\bibinfo {author} {\bibfnamefont {R.}~\bibnamefont
  {Mandal}}\ and\ \bibinfo {author} {\bibfnamefont {P.}~\bibnamefont
  {Sollich}},\ }\href {\doibase 10.1103/PhysRevLett.125.218001} {\bibfield
  {journal} {\bibinfo  {journal} {Phys. Rev. Lett.}\ }\textbf {\bibinfo
  {volume} {125}},\ \bibinfo {pages} {218001} (\bibinfo {year}
  {2020})}\BibitemShut {NoStop}%
\bibitem [{\citenamefont {Parley}\ \emph {et~al.}(2022)\citenamefont {Parley},
  \citenamefont {Sastry},\ and\ \citenamefont
  {Sollich}}]{PhysRevLett.128.198001}%
  \BibitemOpen
  \bibfield  {author} {\bibinfo {author} {\bibfnamefont {J.~T.}\ \bibnamefont
  {Parley}}, \bibinfo {author} {\bibfnamefont {S.}~\bibnamefont {Sastry}}, \
  and\ \bibinfo {author} {\bibfnamefont {P.}~\bibnamefont {Sollich}},\ }\href
  {\doibase 10.1103/PhysRevLett.128.198001} {\bibfield  {journal} {\bibinfo
  {journal} {Phys. Rev. Lett.}\ }\textbf {\bibinfo {volume} {128}},\ \bibinfo
  {pages} {198001} (\bibinfo {year} {2022})}\BibitemShut {NoStop}%
\end{thebibliography}%

%\end{document}

\clearpage

\begin{widetext}

\section*{\large Appendix}

%In this document we provide supplemental figures and details related to the results presented in the main text.

\maketitle

%\clearpage

\setcounter{figure}{0}

\makeatletter 
\renewcommand{\thefigure}{A\@arabic\c@figure}
\makeatother

%\section{SI Fig. S8}
\begin{figure}[h]
	\includegraphics[scale=0.4]{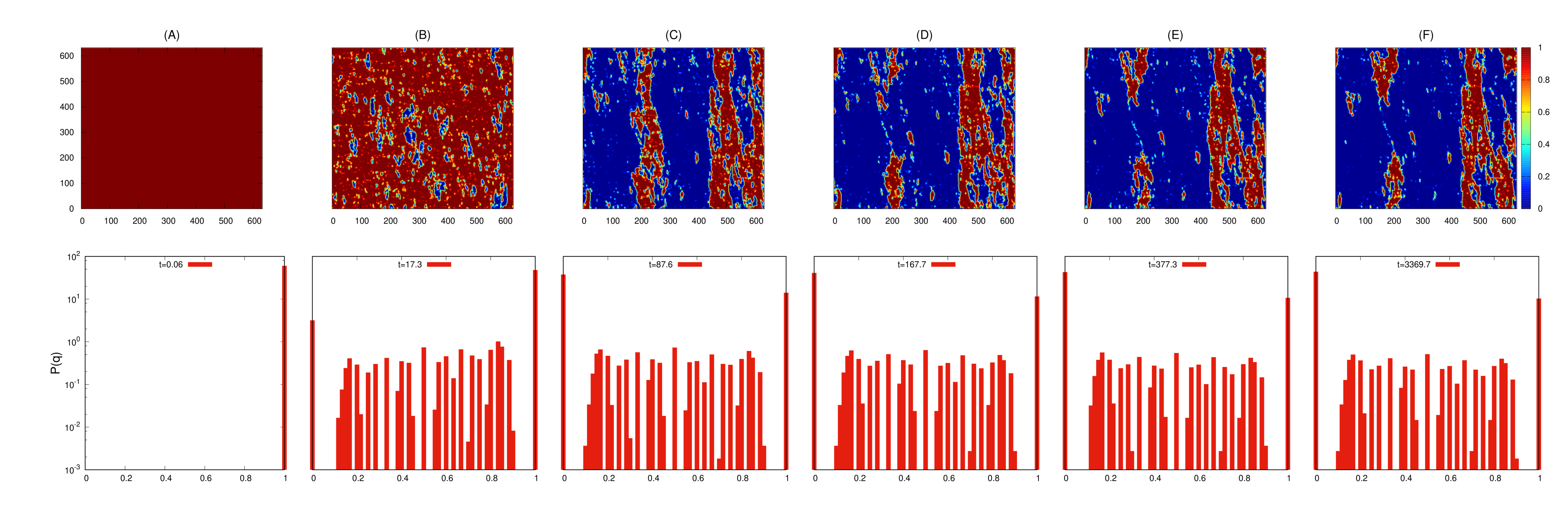}
	\caption{(Top) Maps of coarse-grained overlap function, $Q_\ell(X,Y;t)$ for $\sigma_0=0.79$, computed at  $t= 0.06$ (A), $87.6$ (B), $167.7$ (C), $377.3$ (D) and $3369.7$ (E), using $\ell=4.94$. (Bottom) Distribution of coarse-grained overlap at corresponding times.}
	\label{dif3}
\end{figure}

%\section{SI Fig. S9}
\begin{figure}[h]
	\includegraphics[scale=0.4]{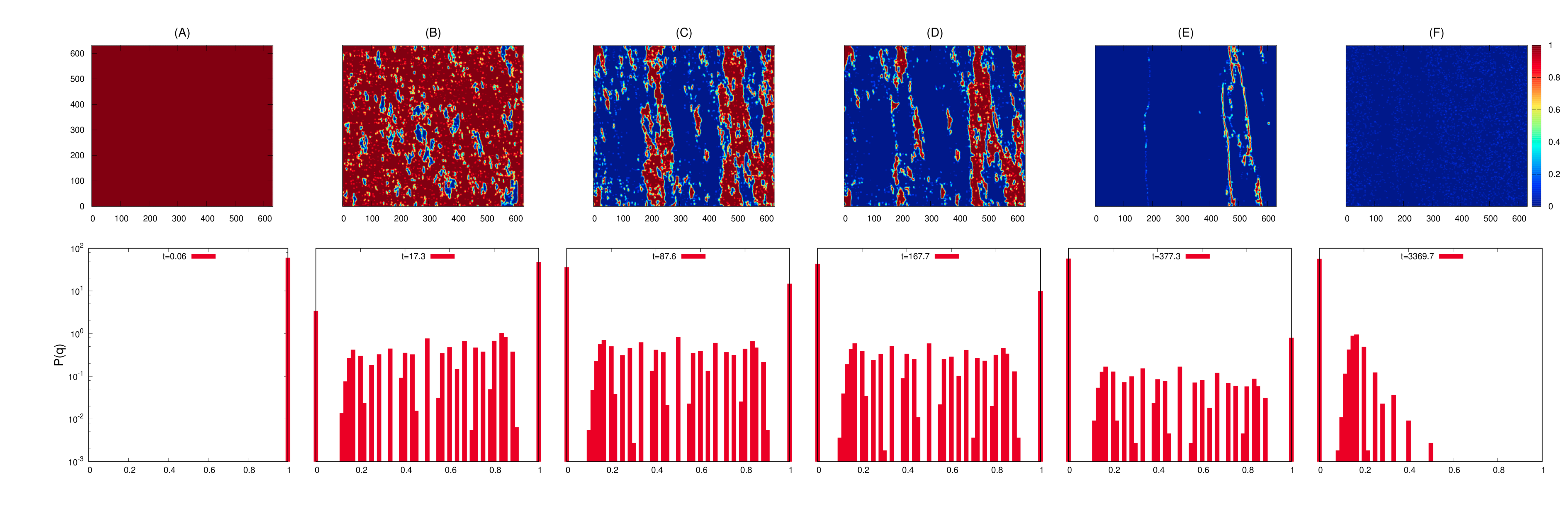}
	\caption{Maps of coarse-grained overlap function, $Q_\ell(X,Y;t)$ for $\sigma_0=0.795$ computed at  $t= 0.06$ (A), $87.6$ (B), $167.7$ (C), $377.3$ (D) and $3369.7$ (E), using $\ell=4.94$. (Bottom) Distribution of coarse-grained overlap at corresponding times.}
	\label{dif3a}
\end{figure}

\newpage

\begin{figure}[h]
\includegraphics[scale=0.15]{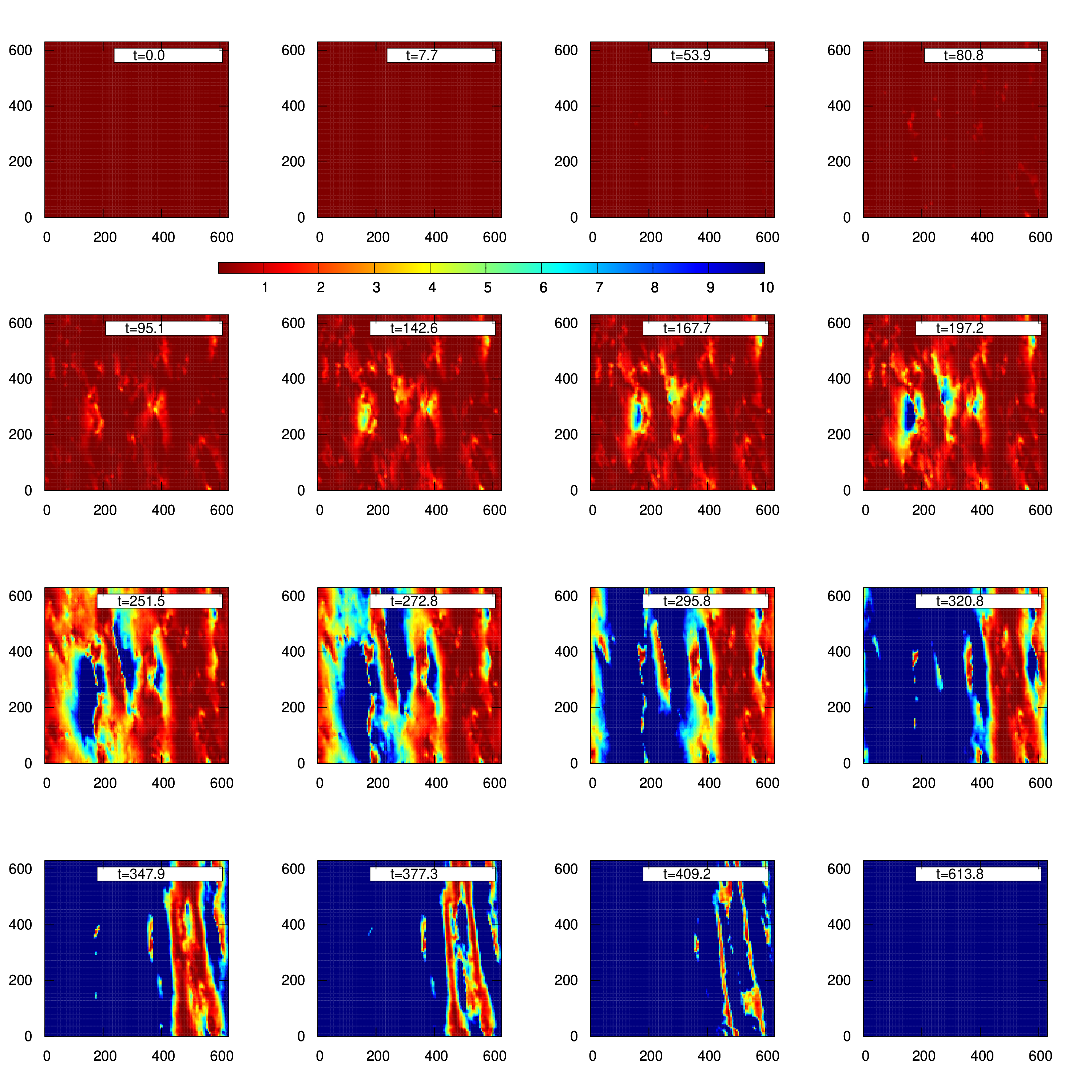}
\caption{Spatial maps of difference in activity (quantified via coarse-grained local MSDs), computed for responses to $\sigma=0.79$ and $\sigma=0.795$, using $\ell=4.94$, starting from same initial condition.}
\label{dif2}
\end{figure}

%\newpage

\end{widetext}

\clearpage

%\bibliography{biblo}

\end{document}